\documentclass[11pt,preprintnumbers,aps,amssymb,nofootinbib,amsmath,superscriptaddress,notitlepage,prd]
{revtex4-2}
\usepackage{epsfig,epsf}
\usepackage{comment}
\usepackage{bm} 
\usepackage{color} 
\usepackage{slashed} 
\usepackage{relsize}	
\usepackage{soul} 
\usepackage{hyperref}
\usepackage{tensor} 
\usepackage{yfonts} 
\newcommand{\beq}{\begin{equation}}
\newcommand{\beql}[1]{\begin{equation}\label{#1}}
\newcommand{\eeq}{\end{equation}}
\def\bal#1\gal{\begin{align}#1\end{align}}
\newcommand{\ball}[1]{\bal\label{#1}}
\newcommand{\eq}[1]{(\ref{#1})}
\newcommand{\fig}[1]{Fig.~\ref{#1}}
\renewcommand{\sec}[1]{Sec.~\ref{#1}}
\DeclareMathOperator{\real}{\mathrm{Re}}

\renewcommand{\b}[1]{{\bm #1}} 
\newcommand{\unit}[1]{\hat {{\bm #1}}} 

\newcommand{\arctanh}{\text{arctanh\,}}
\usepackage{feynmp-auto}

\begin{document}


\title{Electromagnetic bremsstrahlung and energy loss in chiral medium}

\author{Jeremy Hansen}

\author{Kirill Tuchin}

\affiliation{
Department of Physics and Astronomy, Iowa State University, Ames, Iowa, 50011, USA}

\date{\today}

\begin{abstract}
We study electromagnetic radiation by a fast particle carrying electric charge in chiral medium. The medium is homogeneous and isotropic and supports the chiral magnetic current which renders the fermion and photon states unstable. The instability manifests as the chirality-dependent resonances in the bremsstrahlung cross section, which enhance the energy loss in the chiral medium. We compute the corresponding cross sections in the single scattering approximation and derive the energy loss in the high energy approximation.   
 
\end{abstract}

\maketitle


\section{Introduction}

This paper is the third in the series of papers dedicated to the problem of energy loss by a fast particle in the chiral medium due to the electromagnetic interactions. The pivotal feature of chiral media is the peculiar response to the external electromagnetic field caused by the chiral anomaly  \cite{Adler:1969gk,Bell:1969ts}. In homogeneous media, it is given by the chiral magnetic current $\b j=b_0\b B$ \cite{Kharzeev:2004ey,Kharzeev:2007tn,Fukushima:2008xe,Kharzeev:2009fn,Kharzeev:2007jp}, where the chiral magnetic conductivity $b_0$ is assumed to  be a constant. It was discussed  in the context of Weyl and Dirac semimetals \cite{Klinkhamer:2004hg,Li:2014bha,Sukhachov:2018uuz}, the quark-gluon plasma \cite{Zhitnitsky:2012ej,Kharzeev:2015znc} and the axion phenomenology  \cite{Sikivie:2020zpn}.
In our first paper \cite{Hansen:2020irw} we computed the collisional energy loss and found that at high energies it is dominated by the chiral Cherenkov radiation. In the second paper \cite{Hansen:2022nbs} we considered the part of the radiative energy loss that is driven by the resonance at $\b q^2=b_0^2$ in the spatial part of  the photon propagator $D_{ij}(q)$. The corresponding contribution to the bremsstrahlung cross section is proportional to the magnetic moment of the scatterer because $D_{ij}$ couple only to the spatial part of the source current whose leading multipole term is the magnetic moment $\textgoth{M}$. It will be further referred to as the ``magnetic channel".
The emergence of the resonance is a signature of the chiral magnetic instability \cite{Carroll:1989vb,Lehnert:2004hq,Tuchin:2018sqe}. Another manifestation of this instability is the appearance of the chiral run-away modes in the modified photon dispersion relation. At high energies   the anomalous term in the photon dispersion relation is small and can be neglected compared to the main contribution stemming from the pole in $D_{ij}$. This is to say that the resonance in the magnetic bremsstrahlung cross section is weakly dependent on the radiated photon kinematics.

In this article we consider the photon bremsstrahlung in the chiral medium due to the Coulomb part $D_{00}(q)$ of the photon propagator. We dub it the ``electric channel". Although $D_{00}(q)$ itself has no conspicuous dependence on $b_0$, the resonance does emerge in the photon propagator due to the anomaly in the photon dispersion relation. It causes the momentum transfer $\b q^2$ to vanish at finite photon emission angles $\theta$. Thus, unlike the magnetic channel considered in \cite{Hansen:2022nbs}, the 
electric channel is driven exclusively by the anomaly in the photon dispersion relation. Along with the photon propagator, the fermion one also becomes resonant, indicating instability of fermions in the chiral medium \cite{Lehnert:2004hq,Tuchin:2018sqe}. The main goal of this paper is to compute the photon bremsstrahlung cross section in the chiral medium due to the Coulomb term in the photon propagator. 

The paper is structured as follows. In \sec{sec:x-sec} we derive the general expression for the scattering cross section. In \sec{resonance} we observe that the cross section is divergent due to the  resonances in the photon and fermion propagators. The divergences are regulated by the finite width of the unstable modes which is proportional to the fermion relaxation rate $\tau^{-1}$. 
Significantly, the same parameter regulates the divergences in the fermion and photon propagators in the electric channel (but not in the magnetic one discussed in \cite{Hansen:2022nbs}). The final expression for the scattering cross section is quite bulky and is given in Appendix~A. The high energy limit, relevant in most applications, is developed in \sec{high energy} where the resonant structures become very clear. Particularly simple expressions for the cross section are derived at low and high temperatures. Our main results are Eqs.~\eq{sleft}, \eq{sright2} and \eq{g5},\eq{g7} for the bremsstrahlung cross section in various limits. The results for the corresponding energy loss are given in \sec{sec:loss}.
The summary and outlook are presented in  \sec{sec:summary}.

\section{scattering cross section}\label{sec:x-sec}

\begin{figure}[t]
\includegraphics[height=3cm]{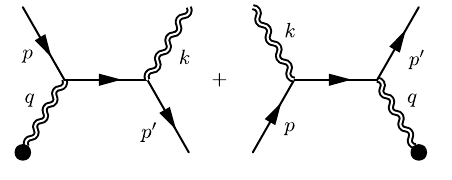} 
  \caption{The two diagrams corresponding to the matrix element for the scattering cross-section. Double lines indicate  photons in chiral medium.}\label{fig2}
\end{figure}

Following the original Bethe and Heitler calculation \cite{Bethe:1934za}, we consider scattering of a charged fermion off a heavy nucleus of mass $M$, electric charge $eZ$ and magnetic moment $\textgoth{M}$. The differential cross section for photon radiation is given by the familiar expression
\ball{c1}
    d\sigma=\frac{1}{2}\sum_{ss'\lambda}\left|\mathcal{M}\right|^2
    \frac{1}{8(2\pi)^5}\frac{\omega |\b p'|}{|\b p|} d\Omega_{\bm{k}} d\Omega' d\omega\,,
\gal
where  $p=(E,\b p)$, $p'=(E',\b p')$ and $k=(\omega,\b k)$ are the incoming fermion, outgoing fermion, and radiated photon momenta respectively. The matrix element $\mathcal{M}$ is represented by two Feynman diagrams shown in \fig{fig2}. 
The corresponding analytical expression reads:
\ball{c3}
    \mathcal{M}=e^2 \overline{u}(p')\left(\slashed{e}_{k\lambda}^*\frac{ \slashed{p}'+\slashed{k}+m}{2p'\cdot k+k^2}\slashed{A}(\b q)-\slashed{A} (\bm{q})\frac{ \slashed{p}-\slashed{k}+m}{2p\cdot k-k^2}\slashed{e}_{k\lambda}^*\right)u(p)\,,
\gal
where $A(\b q)$ indicates the external field, $q=p'-p+k$ is the momentum transfer, $m$ is the mass of the fermion, and $e_{k\lambda}$ is the photon polarization vector. Whereas in vacuum the photon four-momentum is light-like, in chiral media it is not. Moreover, in the Lorentz gauge, photon is forced to be in one of the two  circularly polarized states. This eliminates the residual gauge invariance that is reflected in the Ward identity \cite{Hansen:2022nbs}.
The photon dispersion relation is 
\ball{c4}
    \omega^2=\b k^2+k^2=\b k^2-\lambda b_0 |\b k|\,,
\gal
where $\lambda=\pm 1$ is right or left photon polarization.  

The electromagnetic potential induced by the the electric current $J^\mu$, associated with the nucleus, in a chiral medium is $A_\mu=-iD_{\mu\nu}J^\nu$. The photon propagator in chiral medium in the Lorentz/Landau gauge take form \cite{Hansen:2022nbs}: 
\ball{a5}
D_{\mu\nu}(q)= -i \frac{q^2 g_{\mu\nu}+i\epsilon_{\mu\nu\rho \sigma}b^\rho q^\sigma+b_\mu b_\nu}{q^4+b^2 q^2-(b\cdot q)^2}+i\frac{\left[q^2-(b\cdot q)^2/q^2\right]q_\mu q_\nu + b\cdot q(b_\mu q_\nu+b_\nu q_\mu)}{q^2\left[q^4+b^2 q^2-(b\cdot q)^2\right]}\,,
\gal
where $b_\mu=(b_0,\b 0)$. In the static limit $q^0=0$ the components of the photon propagator read \cite{Qiu:2016hzd}
\begin{subequations}\label{b12}
\bal
&D_{00}(\b q)= \frac{i}{\b q^2}\,,\label{b12a}\\
&D_{0i}(\b q)= D_{0i}(\b q)= 0\,,\label{b12b}\\
&D_{ij}(\b q)=-\frac{i\delta_{ij}}{\b q^2-b_0^2}-\frac{\epsilon_{ijk}q^k}{b_0(\b q^2-b_0^2)}+\frac{\epsilon_{ijk}q^k}{b_0\b q^2}+\frac{iq_iq_j}{\b q^2(\b q^2-b_0^2)}\,.\label{b12c}
\gal
\end{subequations}
The gauge-dependent terms proportional to $q_\mu$ and $q_\nu$ vanish when substituted into the scattering amplitude. The spatial components \eq{b12c} couple to the nucleus magnetic moment and have a resonance at $\b q^2=b_0^2$. 
We analysed this magnetic channel in our previous paper \cite{Hansen:2022nbs}.   
In this paper we are interested in the monopole component of the external field which is determined by the nuclear electric charge. Convolution of the current $J^\nu(\b x) =eZ \delta\indices{^\nu_0}\delta(\b x)$ with the $D_{00}$ component of the photon propagator gives rise to the Coulomb potential:
\ball{b15}
A^0(\b q)= eZ/\b q^2\,,\qquad \b A(\b q)=0\,.
\gal
Plugging  \eq{b15} into \eq{c3} and averaging over the fermion spin directions, we can obtain a fairly compact expression for the  differential cross section of the electric channel:
\ball{matrix}
    d\sigma=\frac{Z^2}{2}\sum_{\lambda}
    \frac{d\Omega_{\bm{k}} d\Omega' d\omega}{4(2\pi)^5}\frac{ |\b p'|}{|\b p|}   \frac{e^6}{\omega\b q^4}
    \Bigg\{4\left|\frac{E p'\cdot e_\lambda}{\kappa'+\frac{k^2}{2\omega}}-\frac{E'p\cdot e_\lambda}{\kappa-\frac{k^2}{2\omega}}\right|^2-\b q^2\left|\frac{p'\cdot e_\lambda}{\kappa'+\frac{k^2}{2\omega}}-\frac{p\cdot e_\lambda}{\kappa-\frac{k^2}{2\omega}}\right|^2\nonumber\\
    +\frac{\omega^2(\b q^2-(\kappa-\kappa'-\frac{k^2}{\omega})^2)}{(\kappa-\frac{k^2}{2\omega})(\kappa'+\frac{k^2}{2\omega})}\nonumber\\
    +\frac{k^2}{4}\Bigg[\b q^2\left(\frac{1}{\kappa-\frac{k^2}{2\omega}}-\frac{1}{\kappa'+\frac{k^2}{2\omega}}\right)^2+\frac{4|p\cdot e_\lambda-p'\cdot e_\lambda|^2}{(\kappa-\frac{k^2}{2\omega})(\kappa'+\frac{k^2}{2\omega})}
    -4\left(\frac{E'}{\kappa-\frac{k^2}{2\omega}}-\frac{E}{\kappa'+\frac{k^2}{2\omega}}\right)^2\Bigg]\Bigg\}\,,
\gal
where $\kappa=p\cdot k/\omega$ and $\kappa'=p'\cdot k/\omega$. In \eq{matrix} the sum over $\lambda$  is left explicit in order to isolate individual  photon polarizations. Eq.~\eq{matrix} reduces to the Bether-Heitler formula in the limit $|\b k|\rightarrow\omega$, or, equivalently,  $k^2\rightarrow 0$.

The photon plane waves can only have a circular polarization which satisfies the transversality condition $k\cdot e_k=\b k\cdot \b e_\lambda=0$ in the Lorentz gauge. We seek computing the cross section for each photon polarization. To perform the angular integrals in \eq{matrix} it is advantageous to eliminate the photon polarization vectors in favor of particle momenta. To this end, we introduce a Cartesian reference frame with the $z$-axis pointing in the direction of the photon momentum $\b k$:
\bal\label{eq14}
\b k = (0,0,|\b k|)\,,\quad 
    \b e_\lambda=\frac{1}{\sqrt{2}}(1,i\lambda ,0)\,,\quad 
    \b p=(p_\perp,0,p_\parallel)\,,\quad 
    \b p'=(p'_\perp\cos{\phi},p'_\perp\sin{\phi},p'_\parallel)\,,
\gal
where $\phi$ is the azimuthal angle. We derive the following identities:
\begin{subequations}%
\bal
    |\b p \cdot \b e_\lambda|^2&= \frac{|\b p\times \b k|^2}{2\b k^2}\,,\label{b100}\\
    |\b p' \cdot \b e_\lambda|^2&= \frac{|\b p'\times\b k|^2(\cos^2\phi+\lambda^2\sin^2\phi)}{2\b k^2}=\frac{|\b p'\times\b k|^2}{2\b k^2}\,,\label{b101}\\
    (\b p\cdot \b e_\lambda)(\b p' \cdot \b e_\lambda^*)&= \frac{|\b p\times\b k||\b p'\times\b k|(\cos\phi-i\lambda\sin\phi)}{2\b k^2}\,,\label{b102}
\gal
\ball{b105}
\b q^2=\b p^2+\b p'^2+\b k^2+2\b k\cdot \b p'-2\b k\cdot \b p-2\frac{\b k\cdot \b p\, \b k\cdot \b p'+|\b k\times\b p||\b k\times\b p'|\cos{\phi}}{\b k^2}\,.
\gal
\end{subequations}
We observe that the dependence on the azimuthal angle $\phi$ appears in \eq{matrix} only by the way of \eq{b102} and \eq{b105}. Therefore, all terms proportional to $\sin\phi$ in \eq{matrix} vanish after integration over the directions of the outgoing fermion. This conclusion holds even after we regulate the divergence in the fermion propagator using \eq{f1} in the next section.
Dropping the imaginary part of  $(\b p\cdot \b e_\lambda)(\b p' \cdot \b e_\lambda^*)$, observing  that
\bal
    p\cdot p'=m^2+\omega(\kappa'-\kappa)+\frac{k^2-q^2}{2}\,,
\gal
and using \eq{eq14},  the non vanishing terms proportional to $\b e_\lambda^i \b e_\lambda^{*j}$ can be written in terms of $\kappa,\kappa'$ and $\b q^2$:
\begin{subequations}\label{polv}
\bal 
 |\b e_\lambda \cdot\bm{ p}|^2=&\frac{|\b k\times\b p|^2}{2\b k^2}=\frac{\b p^2}{2}-\frac{(\b p\cdot \b k)^2}{2\b k^2}
=\frac{\omega^2 E\kappa}{\b k^2}-\frac{k^2}{2\b k^2}E^2-\frac{m^2}{2}-\frac{\omega^2}{2\b k^2}\kappa^2\,,\\
 |\b e_\lambda \cdot\bm{ p'}|^2=&\frac{\omega^2 E'\kappa'}{\b k^2}-\frac{k^2}{2\b k^2}E'^2-\frac{m^2}{2}-\frac{\omega^2}{2\b k^2}\kappa'^2\,,\\
\mathrm{Re}[(\b e_\lambda\cdot\b  p)(\b e_\lambda^*\cdot \b p')]=& \frac{\b p \cdot \b p'}{2}-\frac{(\b p\cdot \b k) (\b p'\cdot \b k)}{2\b k^2}\nonumber\\
=&\frac{\omega^2(\kappa'E'+\kappa E-\kappa\kappa')-\omega k^2(\kappa-\kappa')-k^2EE'}{2\b k^2}-\frac{m^2}{2}-\frac{k^2+\b q^2}{4}\,.
\gal 
 \end{subequations}
 Eqs.~\eq{polv} can then be used to rewrite \eq{matrix} without an explicit dependence on the photon polarization vectors. 

\section{Regulation of the Resonances}\label{resonance}

An examination of the fermion and photon propagators used in the calculation of the differential cross section reveals divergences in three kinematic regions: $\b q^2=0$, $2\omega\kappa=k^2$ and  $2\omega \kappa'=-k^2$. The first one is the familiar infrared Coulomb pole $\b q^2=0$ which is regulated in the usual way 
\ball{f2}
\frac{1}{\b q^2}\to \frac{1}{\b q^2+\mu^2}\,,
\gal
where $\mu$ is the Debye mass of the medium. If the  scattering particle mass $m$ is much larger than $\mu$, the minimum momentum transfer is of the order $m$. However, the Debye mass scales with the temperature and becomes much larger than $m$ in a sufficiently hot medium. Therefore, in the following analysis it will be convenient to consider two limiting cases depending on the relative magnitude of $m$ and $\mu$. In the next section we will revisit the behavior of the photon propagator at small momentum transfers and show that due to the anomaly, $\b q^2$ can vanish at finite $m$. We will revise the regularization procedure accordingly.

The other two divergences occur in the fermion propagator and  reflect its instability in chiral matter with respect to spontaneous photon emission  \cite{Carroll:1989vb,Joyce:1997uy,Boyarsky:2011uy,Kharzeev:2013ffa,Khaidukov:2013sja,Kirilin:2013fqa,Akamatsu:2013pjd,Avdoshkin:2014gpa,Dvornikov:2014uza,Tuchin:2014iua,Manuel:2015zpa,Buividovich:2015jfa,Sigl:2015xva,Xia:2016any,Kaplan:2016drz,Kirilin:2017tdh,Tuchin:2018sqe,Mace:2019cqo}. With the account of the finite width the fermion propagators in \eq{matrix} are modified as follows:  
\ball{f1}
\frac{1}{2\omega\kappa-k^2}\to \frac{1}{2\omega\kappa-k^2+iE/\tau}\,,\qquad \frac{1}{2\omega\kappa'+k^2}\to \frac{1}{2\omega\kappa'+k^2-iE'/\tau}\,,
\gal
where $\tau$ is the relaxation time as measured in the medium rest frame \footnote{In the fermion rest frame the relaxation rate is $(E/m)\tau^{-1}$.}. A number of inelastic processes contribute to the relaxation of the chiral state of the fermion. Among them is the spontaneous photon emission, also known in the literature as the vacuum or chiral Cherenkov radiation. 
Its rate  at the leading order in the perturbation theory is:
\ball{decay}
   W= \frac{e^2}{8\pi}\int \frac{d\omega}{\omega}\left[b_0\left(\frac{\omega^2}{2E^2}-\frac{\omega}{E}+1\right)-\frac{m^2\omega}{ E^2}\right]
   \Theta\left(\omega^*-\omega\right)
   \Theta(\lambda b_0)\,,
\gal
where the threshold photon energy is 
\ball{f5}
\omega^*=\frac{\lambda b_0 E^2}{\lambda b_0 E+m^2}\,
\gal
and $\Theta$ is the step-function \cite{Tuchin:2018sqe}. In particular, when $m^2\gg \lambda b_0 E$, the fermion decay rate is $W\approx \alpha b_0/2$. One can take this as the low bound of the total relaxation rate: $\tau^{-1}> \alpha b_0/2$.

We now have all necessary  ingredients to complete our calculation of the differential cross section. Eq.~\eq{matrix} may be integrated over directions term by term while making substitutions \eq{f1},\eq{f2} in order to obtain the frequency dependence of  the differential cross section. The result is given by \eq{matrix2}, \eq{A0} in Appendix~\ref{sec:Appendix1}. We use it for the numerical calculation presented in \fig{fig:spectrum}.


\section{high energy limit}\label{high energy}

The exact expression for the cross section \eq{matrix2}  is rather complicated.  In applications one is usually interested in the high energy limit $ E, E'\gg m,\mu$ where
the expression for the differential cross section and squared matrix element significantly simplify. To derive the high energy limit, it is convenient to expand the squared amplitude before the integration over the fermion directions: 
\ball{matrel}
    \frac{1}{2}\sum_{s,s'}|\mathcal{M}|^2&\approx\frac{2Z^2e^6}{\omega^2(\b q^2+\mu^2)^2}\real\Bigg[4\left|\frac{ E p'\cdot e_\lambda}{\kappa'+\frac{k^2}{2\omega}+i\frac{ E'}{\tau \omega}}-\frac{ E'p\cdot e_\lambda}{\kappa-\frac{k^2}{2\omega}+i\frac{ E}{\tau \omega}}\right|^2
    \nonumber\\
    &+\frac{\omega^2(\b q^2-(\kappa-\kappa'-\frac{k^2}{\omega})^2)}{(\kappa-\frac{k^2}{2\omega}+i\frac{ E}{\tau \omega})(\kappa'+\frac{k^2}{2\omega}-i\frac{ E'}{\tau \omega})}
    \nonumber\\
    &+k^2\Bigg(\frac{\b q^2}{4}\left|\frac{1}{\kappa-\frac{k^2}{2\omega}+i\frac{ E}{\tau \omega}}-\frac{1}{\kappa'+\frac{k^2}{2\omega}+i\frac{ E'}{\tau \omega}}\right|^2
    +\frac{|p\cdot e_\lambda-p'\cdot e_\lambda|^2}{(\kappa-\frac{k^2}{2\omega}+i\frac{ E}{\tau \omega})(\kappa'+\frac{k^2}{2\omega}-i\frac{ E'}{\tau \omega})}
    \nonumber\\
    &-\left|\frac{ E'}{\kappa-\frac{k^2}{2\omega}+i\frac{ E}{\tau \omega}}-\frac{ E}{\kappa'+\frac{k^2}{2\omega}+i\frac{ E'}{\tau \omega}}\right|^2\Bigg)\Bigg]\,,
\gal
Additionally, we are interested in photon energies $\omega\gg b_0$ which allows us to write the dispersion relation \eq{c4} as 
\ball{h1}
|\b k|\approx \omega+ \frac{\lambda b_0}{2}\,,
\gal
and treat $k^2/
\omega^2\approx -\lambda b_0/ \omega$ as a small parameter. In this limit Eq.~\eq{polv} may be rewritten as
 \begin{subequations}\label{h2}
\bal 
& |\b e_\lambda \cdot\bm{ p}|^2\approx\left(1-\lambda\frac{b_0}{\omega}\right)^2 E\kappa+\left(\frac{\lambda b_0}{2\omega}-\frac{b_0^2}{2\omega^2}\right) E^2-\frac{m^2+(1-\lambda\frac{b_0}{\omega})^2\kappa^2}{2}\,,\\
&  |\b e_\lambda \cdot\bm{ p'}|^2\approx\left(1-\lambda\frac{b_0}{\omega}\right)^2 E'\kappa'+\left(\frac{\lambda b_0}{2\omega}-\frac{b_0^2}{2\omega^2}\right) E'^2-\frac{m^2+(1-\lambda\frac{b_0}{\omega})^2\kappa'^2}{2}\,,\\
& \mathrm{Re}[(\b e_\lambda\cdot\b  p)(\b e_\lambda^*\cdot \b p')]=\frac{(1-\lambda\frac{b_0}{\omega})^2(\kappa' E'+\kappa E-\kappa\kappa')-m^2}{2}+\left(\frac{\lambda b_0}{2\omega}-\frac{b_0^2}{2\omega^2}\right) E E'+\frac{\lambda b_0 \omega-\b q^2}{4}\,.
\gal    
 \end{subequations}
Employing Eqs.~\eq{h2} to get rid of the photon polarization vectors in favor of the momenta in \eq{matrel} we derive 
 \ball{h3}
    \frac{1}{2}\sum_{s,s'}|\mathcal{M}|^2\approx\frac{2Z^2e^6}{\omega^2(\b q^2+\mu^2)^2}&\real
    \Bigg\{\frac{(\b q^2+\mu^2)( E^2+ E'^2+4\frac{\lambda b_0}{\omega} E E')+4 b_0^2  E E'}{(\kappa'+\frac{\lambda b_0}{2}+i\frac{ E'}{\tau \omega})(\kappa-\frac{\lambda b_0}{2}-i\frac{ E}{\tau \omega})}
    \nonumber\\
    &-2\omega^2\left(\frac{\kappa'+\frac{\lambda b_0}{2}+i\frac{ E'}{\tau \omega}}{\kappa-\frac{\lambda b_0}{2}-i\frac{ E}{\tau \omega}}+\frac{\kappa-\frac{\lambda b_0}{2}-i\frac{ E}{\tau \omega}}{\kappa'+\frac{\lambda b_0}{2}+i\frac{ E'}{\tau \omega}}\right)
    \nonumber\\
    &-4\left[m^2-\frac{\lambda b_0}{\omega}( E E'-m^2+\omega^2)\right]\left(\frac{ E'}{\kappa-\frac{\lambda b_0}{2}-i\frac{ E}{\tau \omega}}-\frac{ E}{\kappa'+\frac{\lambda b_0}{2}+i\frac{ E'}{\tau \omega}}\right)^2
    \nonumber\\
    &+\frac{\lambda 4 b_0}{\omega}\left(\frac{( E+ E') E'}{\kappa-\frac{\lambda b_0}{2}+i\frac{ E}{\tau \omega}}-\frac{( E+ E') E}{\kappa'+\frac{\lambda b_0}{2}+i\frac{ E'}{\tau \omega}}\right)\Bigg\}\,.
\gal

The largest contributions come from the resonances at   $2\omega\kappa-k^2=0$ and $2\omega\kappa'+k^2=0$. In the high energy approximation these equations have a solution only for $m\omega-b_0  E E'\leq 0$. This inequality can be  equivalently expressed as the requirements  $\omega\le \omega^*$ and $b_0\lambda>0$ which is consistent with  \eq{decay} and \eq{f5}. Evidently, only one of the two photon polarizations (the one with $b_0\lambda>0$) is resonant. As such, the differing polarization cases must be treated separately.
Consider, for example the pole at  
\ball{j1}
k^2=2\omega\kappa=2p\cdot k\approx \omega E\left(\frac{m^2}{E^2}+\frac{k^2}{\omega^2}+\theta^2\right)\,,
\gal
where $\theta$ is the angle between $\b k$ and $\b p$. 
The denominator of the corresponding fermion propagator is
\ball{j2.1}
\frac{1}{2p\cdot k-k^2+iE/\tau}=\frac{1}{\omega E\left(\frac{m^2}{E\omega}\,\frac{\omega-\omega^*}{E-\omega^*}+\theta^2 + \frac{i}{\omega\tau}\right)}\,,
\gal
where we used \eq{f5} to eliminate $\lambda b_0$ in favor of $\omega^*$:
\ball{j2.5}
k^2\approx -\lambda b_0\omega = -\frac{\omega\omega^* m^2}{E(E-\omega^*)}\,,\quad (\lambda b_0>0)\,.
\gal
Note that \eq{j2.5} makes sense only when $\lambda b_0>0$, for otherwise $\omega^*$ is negative, indicating that there is no instability when $\lambda b_0<0$.
Eq.~\eq{j2.1} implies that above the threshold, viz.\ when $\omega>\omega^*$,
photon is radiated mostly at angles $\theta\lesssim \theta_0=\sqrt{\frac{m^2}{E\omega}\,\frac{\omega-\omega^*}{E-\omega^*}}$. However, at and below the threshold, the angular distribution diverges at 
$\theta= \sqrt{\frac{m^2}{E\omega}\,\frac{\omega^*-\omega}{E-\omega^*}}$ which is regulated by the cutoff introduced in the previous section.

Let us now examine the photon propagator in the resonant case $\lambda b_0>0$. Writing $\b q^2= [(\b k\times \b q)^2+ (\b k\cdot \b q)^2]/\b k^2$ and expanding at small photon emission angles we obtain 
\ball{j3}
\b q^2=-(k+p'-p)^2\approx  \theta^2E^2+\theta'^2E'^2-2EE'\theta \theta'\cos\phi+\frac{1}{4}\left[\frac{m^2(\omega-\omega^*)}{E'(E-\omega^*)}-E \theta^2+E' \theta'^2\right]^2\,.
\gal
In the non-anomalous case $\omega^*=0$, and the momentum transfer is bounded from below by $\frac{m^4\omega^2}{4 E^2E'^2}$. In contrast, in the presence of the anomaly the momentum transfer is allowed to vanish. We can find the corresponding kinematic region by first observing that the sum of the first three terms and the last term in the r.h.s.\ of \eq{j3} are non-negative and therefore have to vanish independently. The sum of the first three terms vanishes only when $\phi=0$ and $E \theta=E' \theta'$ in which case the momentum transfer reads
\ball{j4}
\b q^2|_{\phi=0,E \theta=E' \theta'}\approx \frac{1}{4}\left[\frac{m^2(\omega-\omega^*)}{E'(E-\omega^*)}+\frac{\omega E}{E'} \theta^2\right]^2 =
\frac{1}{4}\frac{\omega^2 E^2}{E'^2}\left[\frac{m^2(\omega-\omega^*)}{\omega E(E-\omega^*)}+\theta^2\right]^2\,.
\gal
This imbues the photon propagator with the same resonant behavior as the fermion propagator, as can be seen by comparing with \eq{j2.1}. Apparently, we need to regulate the divergence at $\b q^2=0$ by replacing  $\theta^2\to \theta^2+\frac{i}{\omega \tau}$ in \eq{j4}. This is tantamount to the replacement $\b q^2\rightarrow\b q^2+\frac{E^2}{4E'^2\tau^2}$. Along with the Debye mass $\mu$ introduced in \eq{f2} it provides the regulator of the photon propagator at small momentum transfers: 
\ball{j10}
\b q^2\rightarrow\b q^2+\frac{E^2}{4E'^2\tau^2}+\mu^2\,.
\gal

The physics of bremsstrahlung in chiral medium is most transparent in two limiting cases: (i) low temperature $m\gg \mu$ and (ii) high temperature $\mu\gg m$. In the anomaly-free medium  $b_0=0$, the first case reduces to the scattering off a single nucleus. In this case the momentum transfer $\b q^2$ never vanishes. On the contrary, in the presence of anomaly, negative $k^2$ can drive the momentum transfer towards zero for one of the photon polarizations, as we explained. 

\subsection{Low temperature $\mu\ll m$}\label{small mu}

We first consider the low temperature/heavy fermion regime. We also assume, for the sake of simplicity, that $\mu^2\ll 1/\tau^2$ so that $\tau$ regulates both the fermion and the photon propagators. 
In the anomaly-free medium, the typical momentum transfer is of the order $m$ and therefore the scattering cross section is insensitive to the cutoffs $\tau$ and $\mu$. This conclusion is upended in the anomalous medium due to the resonances discussed in \sec{resonance}.

The bremsstrahlung cross section is obtained by substituting \eq{h3} into \eq{c1} and integrating over the fermion and photon directions. The angular integrals are performed in Appendix~\ref{appA}. The results are essentially different for positive and negative values of the parameter $b_0\lambda$. For $b_0\lambda<0$ there are no resonances and the angular integrals simplify greatly. In this case, $\tau^{-1}$ and $\mu$ may be neglected given that all integrals are convergent. In this case the cross section reads: 
\ball{sleft}
    \frac{d\sigma(b_0\lambda<0)}{d\omega}\approx \frac{Z^2e^6 E'}{4(2\pi)^3(m^2\omega-\lambda b_0  E E') E}
    \left(\frac{ E}{ E'}+\frac{ E'}{ E}-\frac{2}{3}\right)\left[\ln\frac{4 E^2 E'^2}{\omega(m^2\omega-\lambda b_0  E E')}-1\right]\,.
\gal
In the anomaly-free medium $b_0= 0$  \eq{sleft} reduces to the well-known Bethe-Heitler expression for the bremsstrahlung cross section on a heavy nucleus
\cite{Bethe:1934za,Berestetskii:1982qgu}:
\ball{k5}
    \frac{d\sigma_\text{BH}}{d\omega}\approx \frac{ Z^2e^6 E'}{4(2\pi)^3m^2\omega E}\left(\frac{ E}{ E'}+\frac{ E'}{ E}-\frac{2}{3}\right)\left(\ln\frac{2 E E'}{m\omega}-\frac{1}{2}\right)\,.
\gal
The effect of the anomaly on this photon polarization  is most significant in the infrared region $\omega\ll b_0EE'/m^2$ where the cross section scales as $\log(1/\omega)$. In contrast, without the anomaly it scales as $(1/\omega)\log(1/\omega)$.
Thus the anomaly tends to suppress emission of photons with $b_0\lambda<0$ polarization. 

The situation is essentially different for the $b_0\lambda>0$ polarziation since the corresponding cross section diverges in the limit $\tau^{-1}\to 0$, i.e.\ at small photon emission angles when $\omega\le \omega^*$. As we explained in the previous section, this divergence occurs concurrently in the fermion and the photon propagators and is regulated by shifting  $\theta^2\to \theta^2+\frac{i}{\omega \tau}$ and similarly for $\theta'$.
Keeping only the essential terms and setting $\mu=0$ we obtain:
\ball{sright2}
    &\frac{d\sigma (b_0\lambda>0)}{d\omega}\approx \frac{Z^2e^6 E'}{4(2\pi)^3 E m^2\omega}\left\{\frac{\left(\frac{ E}{ E'}+\frac{ E'}{ E}-\frac{2}{3}\right)\left(\ln\frac{4 E^2 E'^2}{m^2\omega^2\sqrt{\frac{E^2(\omega^*-\omega)^2}{\omega^2(E-\omega^*)^2}+\frac{4 E^4 E'^2}{m^4\omega^4\tau^2}}}-1\right)}{\sqrt{\frac{E^2(\omega^*-\omega)^2}{\omega^2(E-\omega^*)^2}+\frac{4 E^4}{m^4\omega^2\tau^2}}}\right.
    \nonumber\\
    & +\frac{2m^4 (\omega^*-\omega)\tau^3}{E^3E'^4  (E-\omega^*)}\left[E'^6\arctan\frac{m^2(\omega^*-\omega)\tau}{  E'(E-\omega^*)}+E^6\arctan\frac{m^2(\omega^*-\omega)\tau }{ E (E-\omega^*)}\right]\Theta(\omega^*-\omega)\Bigg\}\,,
\gal
 where $\Theta$ is the step function and we replaced $b_0\lambda$ in favor of  $\omega^*$  using \eq{j2.5}. 

The are two kinds of terms in \eq{sright2}. (i) The first line of \eq{sright2} reduces to the anomaly-free result \eq{k5} in the limit $b_0\to 0$ and $ E/\tau\to 0$. We neglected all terms proportional to $\tau^{-1}$ with exception of those appearing under the radicals where their role is to regulate the divergence at the threshold $\omega=\omega^*$. 
(ii) The second and the third lines of \eq{sright2} represent the most singular resonance contributions, viz.\ the terms that are most divergent at small $\tau^{-1}$. In the limit $b_0\to 0$ the step function can only be satisfied when $\omega\to 0$, hence the anomalous contribution vanishes.

 We stress that the chiral resonance in the fermion propagator contributes only to one of the photon polarizations, namely $b_0\lambda>0$. The other polarization $b_0\lambda<0$ is suppressed. This is in contrast to the lack of a distinction between handedness in the absence of the anomaly. A remarkable feature of the anomalous contribution, dominated by the second and the third lines of \eq{sright2}, is that at small $\omega$ its spectrum scales as $1/\omega$, which is the same form as the soft photon spectrum without the chiral anomaly \eq{k5}. 
 
 \begin{figure}[t]
\begin{tabular}{cc}
      \includegraphics[height=5.5cm]{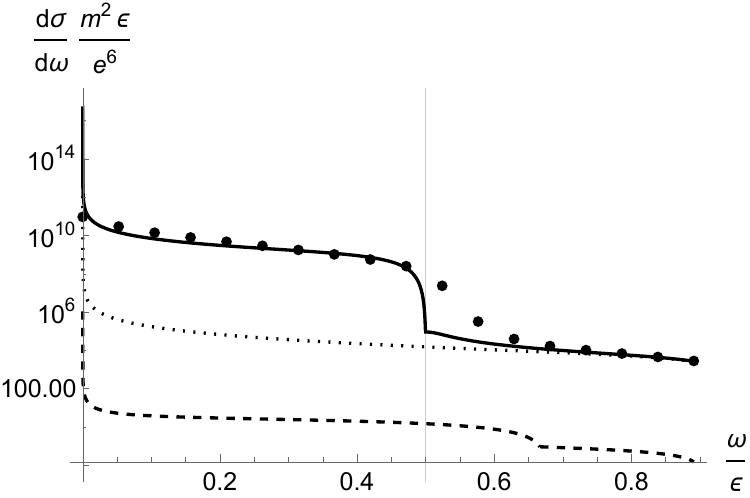} &
      \includegraphics[height=5.5cm]{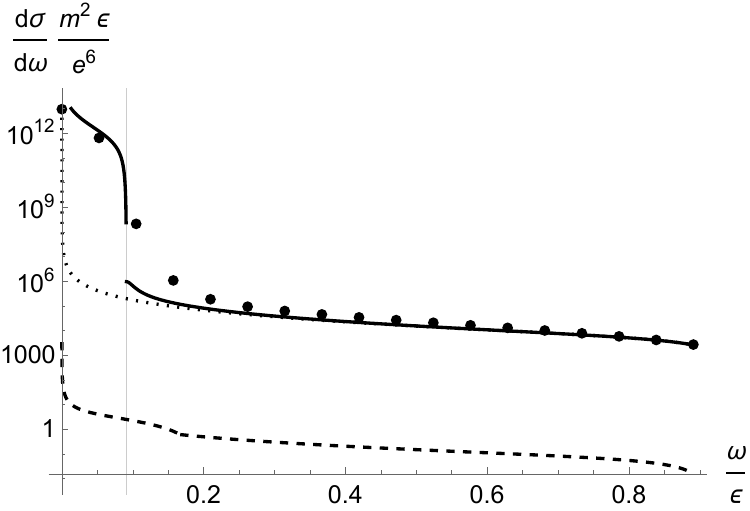}
      \end{tabular}
  \caption{The bremsstrahlung spectrum at $\mu\ll m$. 
  The dots: the exact leading order formula \eq{matrix2}, solid line: the high energy approximation given by the sum of equations \eq{sleft} and \eq{sright2}, dotted line: the Bethe-Heitler expression \eq{k5}, dashed line: the magnetic moment contribution  \eq{j5}. 
Left panel:  $b_0=0.1m$,  $\mu=10^{-3}b_0$, $\tau^{-1}= 0.1b_0$. Right panel: $b_0=10^{-2}m$, $\mu=10^{-3}b_0$, $\tau^{-1}=0.1 b_0$. Both panels: $E=10m$,  $Z=33$, $\textgoth{M}=\mu_N$, $\Gamma=0.1b_0$. The vertical line indicates $\omega^*/E$.}
\label{fig:spectrum}
\end{figure}
 
 The bremsstrahlung photon spectra are plotted in \fig{fig:spectrum}.  Inspection of \fig{fig:spectrum} reveals two significant features. Firstly, at $\omega<\omega^*$ the cross section is enhanced as compared to the Bethe-Heitler formula.  To better understand the parametric dependence of the cross section, consider  the soft photon region  $\omega\ll \omega^*\ll E$, in particular $\omega^* \approx \lambda b_0E^2/m^2$. Eq.~\eq{sright2} can be written as 
\ball{h10}
\frac{d\sigma (b_0\lambda>0)}{d\omega}\approx  \frac{3\pi}{2} \,\frac{m^2\tau^3 b_0}  {\ln\frac{2E^2}{m\omega}}\,\frac{d\sigma _\text{BH}}{d\omega}.
\gal
Since $m\tau \gg b_0\tau\gg 1$ we indeed observe that the anomalous cross section is strongly enhanced in the soft photon region. 
 Secondly, when focusing on the production of the right-handed photons, we see a sharp cut-off at $\omega=\omega^*$. The anomalous contribution vanishes at higher photon energies. A similar behavior was discussed in our paper \cite{Hansen:2022nbs} for the case of magnetic moment contribution.

To complete our discussion, we cite here the magnetic moment contribution to the bremsstrahlung in the semi-soft photon limit $b_0\ll \omega\ll  E$ computed in \cite{Hansen:2022nbs}: 
\ball{j5}
\frac{d\sigma_M}{d\omega}\approx\frac{ 2e^4 \textgoth{M} ^2  }{3(2\pi)^3 \omega } \left[\frac{3b_0^2}{m^2}\ln(\frac{4 E^4}{m^2\omega^2})+\ln^2\frac{4 E^2}{m^2}+\frac{2b_0^4\pi}{m^2\Gamma^2} \Theta(\omega_0-\omega) \right]\,,
\gal
where $\textgoth{M}$ is the magnetic moment. In \fig{fig:spectrum} one can see that it gives a minor correction to the total cross section, even though it exhibits qualitative features similar to \eq{sleft} and \eq{sright2}.  
 Overall the anomaly in the magnetic channel may amplify photon production over a wider range of frequencies but  produces less radiation  when compared to the electric channel.

\subsection{High temperatures $\mu\gg m$}\label{high T}

We now consider the differential cross-section in the opposite limit where $\mu\gg m$ and $\mu^2\gg 1/\tau^2$, so that $\tau^{-1}$ is neglected in the photon propagator.  The process of computing the differential cross-section and the necessary integrals is similar to the one outlined in Appendix~\ref{appA}, except that $m$ may be taken to zero in the high-temperature limit. It is well-known  that $\mu$ not only regulates the Coulomb pole, but also sets the smallest photon emission angle $\theta_\text{min}=\mu/ E\ll 1$ \cite{Baier:1996vi,Baier:1994bd}. In the previous subsection both roles were played by $m$.  In the high energy limit, one can  take the integrals  $I_{j,n,l}$, by expanding the integrands at small photon emission angles $\theta$ and $\theta'$:
\ball{g1}
 I_{j,n,l}\approx &\int\frac{d\Omega'd\Omega_k}{ \left[( E^2\theta^2-2  E E \theta\theta' \cos\phi+ E'^2\theta'^2)+\frac{1}{4}( E\theta^2- E'\theta'^2+\lambda b_0)^2\right]^j}\nonumber\\&\times\frac{1}{\left[( E\omega\theta^2-\lambda b_0  E')^n-(-i \frac{E}{\tau})^n\right]\left[( E'\omega\theta'^2-\lambda b_0  E)^l-(i\frac{E'}{\tau})^l\right]}\,,
\gal
where $\phi$ is the azimuthal angle which may be integrated over directly. As an example consider $I_{1,1,1}$. Integrating of $\phi$ yields:
\ball{g3}
 I_{1,1,1}\approx &\int_{\frac{\mu^2}{ E^2}}^\infty\int_{\frac{\mu^2}{ E'^2}}^\infty\frac{\pi^2d\theta'^2d\theta^2}{ \sqrt{(( E^2\theta^2 + E'^2\theta'^2)+\frac{1}{4}( E\theta^2- E'\theta'^2+\lambda b_0)^2)^2-4  E^2 E'^2 \theta^2\theta'^2}}\nonumber\\&\times\frac{1}{(( E\omega\theta^2-\lambda b_0  E')+i \frac{E}{\tau})(( E'\omega\theta'^2-\lambda b_0  E)-i\frac{E'}{\tau})}\,,
\gal
where we replaced the upper integration limits by infinities thanks to the fast convergence of the integrals. Taking the remaining integrals we find
that $I_{1,1,1}$ gives the following contribution to the differential cross-section 
\ball{g4}
 I_{1,1,1}\approx &\frac{4\pi^2\ln\frac{4 E^2 E'^2}{\sqrt{(\mu^2\omega^2- \lambda b_0 \omega E E')^2+4 \frac{E^4E'^2}{\tau^2}}}}{\omega E E'\sqrt{(\mu^2\omega- \lambda b_0  E E')^2+4 \frac{E^3 E'}{\tau^2}}}\,.
\gal
The other terms may be computed in a similar fashion. In those integrals where the integral over $\theta$ (or $\theta'$) is not convergent at large angles, one needs to first integrate over the region $\theta<\theta'$ (or $\theta'<\theta$); the remaining integral then converges well. The result of the calculation is different for the two photon polarziations and reads 
\ball{g5}
    \frac{d\sigma(b_0\lambda<0)}{d\omega}\approx \frac{Z^2e^6 E'}{4(2\pi)^3(\mu^2\omega -\lambda b_0  E E') E}
    \left(\frac{ E}{ E'}+\frac{ E'}{ E}\right)\left[\ln\frac{4 E^2 E'^2}{\mu^2\omega^2-\lambda b_0 \omega E E'}-1\right]\,,
\gal
\ball{g7}
    &\frac{d\sigma (b_0\lambda>0)}{d\omega}\approx \frac{Z^2 E'e^6}{4(2\pi)^3 E\mu^2\omega}
    \Bigg\{ \frac{\mu^2\omega(\frac{ E}{ E'}+\frac{ E'}{ E})\big[\ln\frac{4 E^2 E'^2}{\sqrt{(\mu^2\omega^2-\lambda b_0 \omega E E')^2+4 \frac{E^4E'^2}{\tau^2}}}-1\big]}{\sqrt{(\mu^2\omega-\lambda b_0  E E')^2+ \frac{E^3E'}{\tau^2}}}
    \nonumber\\
     &  +\frac{\lambda  b_0\mu\tau }{  E} \arctan\left(\frac{(\lambda b_0  E E'-\mu^2\omega)\tau}{ E E'}\right)\Theta(\omega^*-\omega)\bigg\}\,,
\gal
where $\omega^*$ is given by \eq{f5}  and $ E', E\gg \mu$.

Consider the soft photon limit in a medium such that $\mu\gg m\gg \tau^{-1}$. Soft photons region corresponds to  $\omega\ll E$ and $\omega\ll \omega^*$, the latter condition implying $ \omega\ll b_0E^2/m^2$.  Then \eq{g7} reduces to 
\ball{g10}
\frac{d\sigma (b_0\lambda>0)}{d\omega}\approx  \frac{\pi}{2} \frac{\mu}{E}  \frac{b_0\tau}{\ln\frac{2E^2}{\mu\omega}} \frac{d\sigma _\text{BH}}{d\omega}\,,
\gal
where the Bethe-Heitler cross section in this case is obtained by setting $b_0=0$ in \eq{g5}. One observes only a modest enhancement, if any, as compared to \eq{h10}.

\begin{figure}[t]
\begin{tabular}{cc}
      \includegraphics[height=5.5cm]{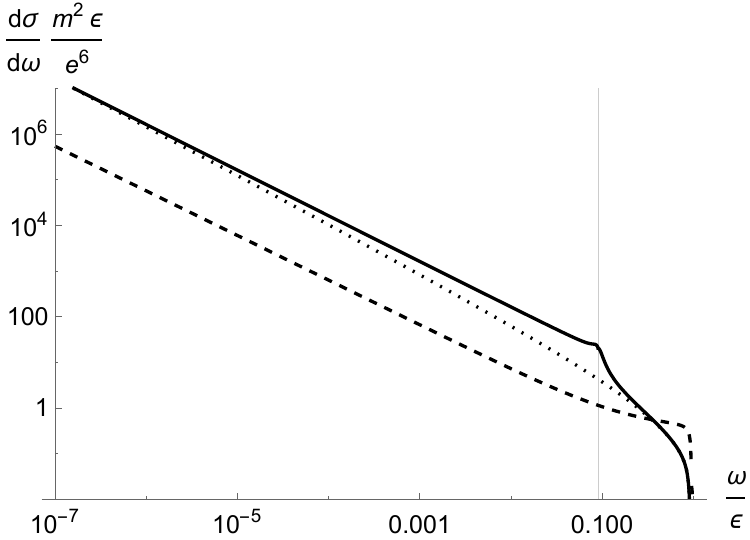} &
      \includegraphics[height=5.5cm]{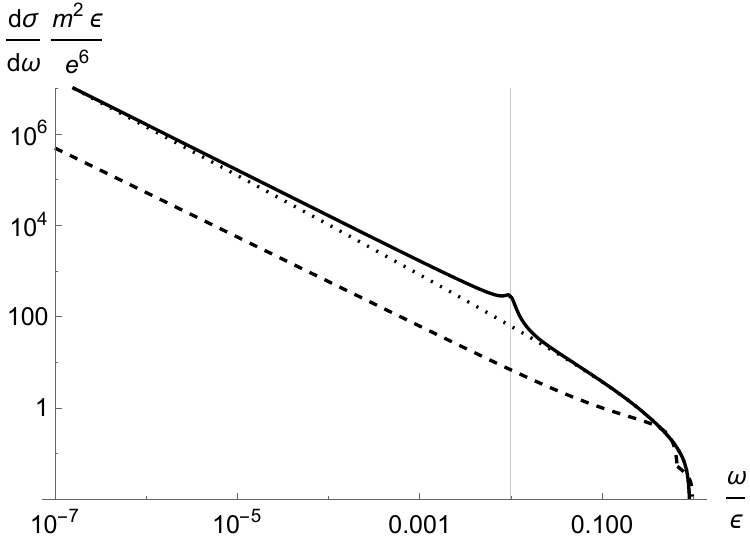} 
      \end{tabular}
  \caption{ The bremsstrahlung spectrum at $\mu\gg m$. Solid line: the high energy approximation given by the sum of equations \eq{g5} and \eq{g7}, dotted line: the Bethe-Heitler limit $b_0=0$, dashed line: the magnetic moment contribution \eq{j5}. Left panel: 
  $b_0=10^{-1} m$, right panel: $b_0=10^{-2}m$, both panels: $E=10^2 m$,  $\mu=10 m$, $\tau^{-1}=\Gamma=0.1 b_0$, $Z=33$. The vertical line indicates $\omega^*/E$.}
\label{fig:spectrum2}
\end{figure}

\section{Energy loss}\label{sec:loss}

A fast particle moving across a medium  losses energy. There are two main mechanisms for energy loss at sufficiently high energies. The collisional energy loss occurs due to momentum transfer in the course of elastic collisions. The radiative energy loss is caused by the bremsstrahlung. In chiral media there is an additional contribution to the energy loss  driven by the chiral anomaly. The results of the previous section allow us to compute the amount of energy lost in a scattering off a single source of electric charge $eZ$. 

Throughout this paper we neglected the quantum interference effects assuming that the photon formation time is much shorter than the mean-free-path $\ell$. In this approximation  
the energy loss per unit length is:
\begin{align}\label{e.loss.1}
-\frac{d E}{d z}= n\int_0^E\omega \frac{d\sigma^{eZ\to eZ\gamma}}{d\omega}d\omega \,,
\end{align}
where $n$ is number of scatterers per unit volume which can be expressed in terms of the elastic scattering cross section $n=1/\ell\sigma^{eZ\to eZ}$. At high energies, 
\bal\label{e.loss.2}
\sigma^{eZ\to eZ}= \frac{\alpha (eZ)^2}{\max\{\mu^2,m^2\}}+\mathcal{O}(\textgoth{M})\,,
\gal
where $\alpha= e^2/4\pi$.
The omitted term in \eq{e.loss.2} is proportional to the small magnetic moment $\textgoth{M}$ of the source. Actually, this term is affected by the anomaly and can become essential at very high energies \cite{ Tuchin:2020gtz}. However, in the present calculation it can be neglected. 

As in the previous section we will be concerned with the two limits.

\subsection{Low temperatures $\mu\ll m$}

Substituting \eq{sleft} or \eq{sright2} into \eq{e.loss.1} and integrating over $\omega$ gives the energy 
loss for two photon polarizations.  For brevity we will record these expressions in the limit $m^2\gg b_0 E$:
\bal\label{e.loss.5}
  -\frac{d E(b_0\lambda<0)}{d z}\approx \frac{e^2 E}{16\pi^2\ell}\left[\ln\frac{2 E}{m}-\frac{1}{3}-\frac{2b_0  E}{9m^2}\left(\pi+2\ln\frac{ E}{b_0}\right)\right]\,,
\gal
\bal\label{e.loss.6}
 -\frac{d E(b_0\lambda>0)}{d z}\approx & \frac{e^2 E}{16\pi^2\ell}
    \Bigg\{\ln\frac{2 E}{m}-\frac{1}{3}+\frac{2b_0  E}{9m^2}\left(\pi+2\ln\frac{ E}{b_0}\right)
     +2\tau   (E-\omega^*)\arctan\frac{2m^2\omega^*\tau}{  E(E-\omega^*)}\Bigg\}\,.
\gal
Eq.~\eq{e.loss.5} agrees with Eq.~(93.24) in \cite{Berestetskii:1982qgu} when $b_0=0$. We observe that the anomaly slightly reduces the amount of energy lost due to the bremsstrahlung of $b_0\lambda<0$ photons. This  is consistent with our discussion in the previous section that the bremsstrahlung cross section is reduced in this case. 

As far as the $b_0\lambda >0$ polarization is concerned, we consider again the soft photon region $\omega\ll \omega^*\ll E$ were $\lambda b_0\ll m$ so that the last term on the r.h.s.\ of \eq{e.loss.6} is dominant. This allows us to cast the energy loss \eq{e.loss.6} in a convenient form:
\ball{e.loss.7}
-\frac{d E(b_0\lambda>0)}{d z}\approx
\frac{\pi\tau E}{\ln \frac{2E}{m}}\left(-\frac{d E}{d z} \right)_\text{BH}\,.
\gal
Thus, the energy loss due to the chiral anomaly is driven by the $b_0\lambda>0$ photon polarization and is enhanced by the large factor $\tau E$ over the Bethe-Heitler result.

\subsection{High temperature $\mu\gg m$}

The rate of energy loss in the high-temperature limit is computed similarly. Plugging \eq{g5} and \eq{g7} into \eq{e.loss.1} we derive
\bal\label{e.loss.14}
    -\frac{d E(b_0\lambda<0)}{d z}&\approx \frac{e^2 E}{16 \pi^2 \ell}\left(\ln\frac{2 E}{\mu}-\frac{6b_0  E}{7\mu^2}\ln\frac{ E}{b_0}\ln\frac{4 E^2}{\mu^2}\right)\,,\\
\label{e.loss.15}
    -\frac{d E(b_0\lambda>0)}{d z}&\approx \frac{e^2 E}{16 \pi^2 \ell}\left(\ln\frac{2 E}{\mu}+\frac{6b_0  E}{7\mu^2}\ln\frac{ E}{b_0}\ln\frac{4 E^2}{\mu^2}+\frac{ 2 b_0 \mu\tau}{ 3 E}\right)\,.
\gal
Eq.~\eq{e.loss.14} agrees with \cite{Baier:1994bd} when $b_0=0$. As in the  low temperature case, the energy loss is mostly driven by $b_0 \lambda>0$ polarization. However, the overall magnitude of the lost energy sensitively depends on the actual values of the parameters.


\section{Summary}\label{sec:summary}

In this paper we computed and analysed the spectrum of electromagnetic radiation emitted by a fast particle  traveling in the chiral medium supporting the chiral magnetic current with constant chiral conductivity $b_0$. We observed in \fig{fig:spectrum} that at low temperatures $\mu\ll m$ the anomalous contribution to bremsstrahlung, stemming from one of the photon polarizations, is several orders of magnitude larger than the non-anomalous one. We also computed the corresponding enhancement of the energy loss. If confirmed by experimental observation, it can serve as an effective tool to study the chiral anomaly in the materials. It can also be used to search for the new forms of the chiral matter. In particular, a cosmic ray moving through the chiral domain generated by an axion field would radiate and lose energy in a peculiar way described in this paper. In the opposite limit of very high temperatures $\mu\gg m$, the effect of the anomaly is much smaller as seen in \fig{fig:spectrum2}.

In arriving at our conclusions we made a number of assumptions. First of all, we treated the  chiral conductivity $b_0$ as a  constant. However, it does evolve in time albeit slowly. Its temporal evolution is characterized by the relaxation time $\tau$, which we assumed to be the large parameter in our calculation. The spatial extent of the domain can be neglected as long as it is larger than photon formation time $t_f$ which is a very good approximation considering that the formation time must be shorter than the mean free path $\ell$ in the Bethe-Heitler limit. Also, for the sake of simplicity we neglected  the chiral displacement $\b b$. The electric current $\b j = \b b\times  \bm{\mathcal{E}}$  is induced at finite $\b b$  and is responsible for the anomalous Hall effect.  At $b_0=0$ it induces the radiative instability of the chiral matter which is  similar in many ways to the homogeneous and isotropic case $b_0\neq 0$, $\b b=0$, especially at high energies \cite{Huang:2018hgk,Tuchin:2018mte}. In fact, one can obtain the rate of the Cherenkov radiation by simply replacing $b_0\to |\b b|\cos\beta$, where $\beta$ is the angle between $\b b$ and $\b k$. The relationship between the two cases is more delicate for bremmstrahlung. The photon propagator takes now the following form, in place of \eq{b12}: 
\bal
&D_{00}= \frac{i\b q^2}{\b q^4+(\b b\times \b q)^2}\,,\label{z6a}\\
& D_{0i}= \frac{(\b b\times \b q)_i}{\b q^4+(\b b\times \b q)^2}\,,\label{z6b}\\
&D_{ij}= 
-\frac{i}{\b q^4+(\b b\times \b q)^2}\left\{
\b q^2 \delta_{ij}+b_ib_j-
\frac{[\b q^4-(\b b\cdot \b q)^2]q_iq_j}{\b q^4}
-\frac{\b b\cdot \b q}{\b q^2}(b_iq_j+b_jq_i)\right\}\,.\label{z6c}
\gal
The denominator of the $D_{00}$ component that drives the electric channel is merely shifted $\mu^2\to \mu^2+(\b b\times \unit q)^2$.
The corresponding energy loss is anisotropic as it depends on the angle between the incident fermion momentum and the displacement vector. Otherwise, we believe that the qualitative similarity will persist. The magnetic channel is quite different, but its contribution to the overall energy loss is not significant. Clearly, due to possible applications to Weyl and Dirac semi-metals, the case of finite $\b b$ deserves a dedicated study. 

Another critical assumption we made is that photon formation time $t_f$ is much shorter than the mean-free path $\ell$. As argued in \cite{Baier:1994bd}, the condition $t_f\ll \ell$ translates into the requirement that  $m,\mu\ll  E\ll \mu\sqrt{\omega\ell}$. Since $t_f$ is proportional to $\omega$, this approximation breaks down at high energies where one must take account of the multiple scatterings of the projectile in the medium. The resulting LPM quantum interference effect \cite{Landau:1953um,Migdal:1956tc} is an important feature of the bremsstrahlung spectrum and must certainly be taken into account at higher energies.  

Throughout this paper, we have treated the chiral anomaly as it pertains to QED. The non-abelian version of the chiral anomaly has a similar effect in QCD generating contributions to gluon and photon bremsstrahlung.  Calculation of bremsstrahlung and energy loss in QCD in the presence of the chiral magnetic current is relevant to the quark-gluon plasma phenomenology. We plan to address this elsewhere.

\acknowledgments
This work  was supported in part by the U.S. Department of Energy Grants No.\ DE-FG02-87ER40371 and No.\ DE-SC0023692.

\appendix

\section{The bremsstrahlung cross section at the leading order}\label{sec:Appendix1}

The bremsstrahlung cross section corresponding to \fig{fig2} including the cutoffs reads:  
\ball{matrix2}
    d\sigma=\sum_\lambda &
    \frac{Z^2\omega e^6 d\omega}{16(2\pi)^5}\frac{ |\b p'|}{|\b p|} \nonumber\\
    &
    \times\real\Bigg\{
    \left(2 E^2+2 E'^2+k^2\right)I_{1,1,1}-\frac{\omega^2+\b k^2}{4\b k^2}(I_{2,1,-1}+I_{2,-1,1})+(I_{1,0,1}-I_{1,1,0})
    \nonumber\\
    &
    +\frac{2m^2(\omega^2+\b k^2)+4 E E'k^2+k^4}{4\b k^2}(I_{1,2,0}-2I_{1,1,1}+I_{1,0,2})\nonumber\\
    &
    -\frac{4}{\b k^2}\big[m^2\b k^2( E'^2I_{2,2,0}-2 E E'I_{2,1,1}+ E^2 I_{2,0,2})+\frac{\b k^2 k^2}{2}\left( E^2I_{2,0,2}+ E'^2I_{2,2,0}\right)\nonumber\\
    &
    + E'^2 E'^2k^2(I_{2,2,0}-2I_{2,1,1}+I_{2,0,2})- E E'k^2(I_{2,0,1}+I_{2,1,0})\big]\nonumber\\
    & 
       +4\pi^2\frac{\arctanh\left(\frac{2|\b k||\b p|}{2\omega E-k^2+i \frac{E}{\tau}}\right)\arctanh\left(\frac{2|\b k||\b p'|}{2\omega E'+k^2-i \frac{E'}{\tau}}\right)}{\b k^2 |\b p||\b p'|}\nonumber\\
    &
    -k^2(4\pi)^2\frac{\arctanh\left(\frac{2\omega E+2|\b p||\b k|-k^2+\mu^2}{\mu |\b p'|}\right)-\arctanh\left(\frac{2\omega E+2|\b p||\b k|-k^2+\mu^2}{\mu |\b p'|}\right)}{\b k^3|\b p||\b p'|\mu}
    \Bigg\}\,,
\gal
where the angular integrals $I_{j,n,l}$ are defined as
\ball{A0} I_{j,n,l}=\int\frac{d\Omega'd\Omega_k}{ (\b q^2+\mu^2)^j\left[(2\omega\kappa-k^2)^n-(-i \frac{E}{\tau})^n\right]\left[(2\omega\kappa'+k^2)^l-(i\frac{E'}{\tau})^l\right]}\,.
\gal
where $j,n,l$ are integers, $k^2=-\lambda b_0|\b k|$. When any of these integers is negative the corresponding cutoffs can be set to zero.

\section{Integrals $I_{j,n,l}$ in the ultrarelativistic Heavy fermion limit}\label{appA}

An analysis of \eq{h3} reveals  several dominant contributions to the differential ultrarelativistic cross-section in terms of $I_{j,n,l}$. We consider these contributions under two different regimes. In the first, we consider the case of heavy fermions relative to the Debye mass such that $\mu\ll m$, $\mu^2\ll E/\tau$,  and in the second we consider the case of high temperature which takes on the opposite limit $m\ll\mu$. We focus on the first case in this appendix, however in both cases we take the high energy limit $\mu,m\ll E, E'$ and $\omega\gg b_0$. Additionally, the dominance of small emission angles allows us to neglect the contribution due to large angles. Letting $\theta$ and $\theta'$ be the angles between $\b k$ and $\b p$, and $\b k$ and $\b p'$ leads to the following approximation
 \ball{App1}
& I_{j,n,l}\approx \int\frac{d\Omega'd\Omega_k}
 {
 \left[\left( E^2\theta^2-2  E' E \theta\theta' \cos\phi+ E'^2\theta'^2\right)+\frac{1}{4}\left(\frac{m^2\omega}{ E E'}- E\theta^2+ E'\theta'^2-\lambda b_0+\frac{2i}{\omega\tau}\right)^2\right]^j 
}\nonumber\\
&\times
\frac{1}{\left[\left( E\omega\theta^2+\frac{m^2\omega-\lambda b_0  E' E}{ E}\right)^n-\left(-i\frac{2E}{\tau} \right)^n\right]\left[\left( E'\omega\theta'^2+\frac{m^2\omega-\lambda b_0  E' E}{ E'}\right)^l-\left(i\frac{2E'}{\tau}\right)^l\right]}\,,
\gal
where $\phi$ is the azimuthal angle ranging from $0$ to $2\pi$. Integrating up to emission angles for the two possible cases $j=1,2$
 \ball{App2}
 &I_{1,n,l}\approx \int\frac{\pi^2d\theta'^2d\theta^2}{ \sqrt{\left[ E^2\theta^2+ E'^2\theta'^2+\frac{1}{4}\left(\frac{m^2\omega}{ E E'}- E\theta^2+ E'\theta'^2-\lambda b_0+\frac{2i}{\omega\tau}\right)^2\right]^2-4  E'^2 E^2 \theta^2\theta'^2}}\nonumber\\&\times\frac{1}{\left[\left( E\omega\theta^2+\frac{m^2\omega-\lambda b_0  E' E}{ E}\right)^n-\left(-i \frac{2E}{\tau}\right)^n\right]\left[\left( E'\omega\theta'^2+\frac{m^2\omega-\lambda b_0  E' E}{ E'}\right)^l-\left(i\frac{2E'}{\tau}\right)^l\right]}\,,
\gal
 \ball{App3}
 &I_{2,n,l}\approx \int\frac{\left[ E^2\theta^2+ E'^2\theta'^2+\frac{1}{4}\left(\frac{m^2\omega}{ E E'}- E\theta^2+ E'\theta'^2-\lambda b_0\right)^2+\mu^2\right]\pi^2d\theta'^2d\theta^2}{ \left\{\left[ E^2\theta^2+ E'^2\theta'^2+\frac{1}{4}\left(\frac{m^2\omega}{ E E'}- E\theta^2+ E'\theta'^2-\lambda b_0+\frac{2i}{\omega\tau}\right)^2\right]^2-4  E'^2 E^2 \theta^2\theta'^2\right\}^\frac{3}{2}}\nonumber\\
 &\times
 \frac{1}{\left[\left( E\omega\theta^2+\frac{m^2\omega-\lambda b_0  E' E}{ E}\right)^n-\left(-i \frac{2E}{\tau}\right)^n\right]\left[\left( E'\omega\theta'^2+\frac{m^2\omega-\lambda b_0  E' E}{ E'}\right)^l-\left(i\frac{2E'}{\tau}\right)^l\right]}\,
\gal

These integrals may then be taken for the relevant factors of $n$ and $l$, noting the dominance of the regions $\omega E^2\theta^2<\sqrt{(m^2\omega-\lambda b_0  E' E)^2+ \frac{E^4}{\tau^2}}$, and $\omega E'^2\theta'^2<\sqrt{(m^2\omega-\lambda b_0  E' E)^2+ \frac{E'^4}{\tau^2}}$, acting as effective bounds for the integration. It is convenient to replace the integrals over emission angles with the difference $\Delta=| E\theta- E'\theta'|$ given that the integrands are largest at $\Delta=0$. The results for various relevant integrals for the differential cross-section are given. For instances
\ball{App4}
 I_{1,1,1}\approx &\frac{4\pi^2\ln\frac{4 E^2 E'^2}{\sqrt{(m^2\omega^2- \lambda b_0 \omega E E')^2+\frac{16 E^4 E'^2}{\tau^2}}}}{\omega E E'\sqrt{(m^2\omega- \lambda b_0  E E')^2+\frac{16 E^3 E'}{\tau}}}\,
\gal
and
\ball{App5}
 I_{1,-1,1}+I_{1,1,-1}\approx &\frac{4\pi^2}{\omega\sqrt{(m^2\omega- \lambda b_0  E E')^2+\frac{16 E^3 E'}{\tau^2}}}\left(\frac{ E'}{ E}+\frac{ E}{ E'}\right)\,.
\gal
Other integrals are more easily taken together, such as 
\ball{App6}
  E'^2I_{2,2,0}&-2 E E'I_{2,1,1}+ E^2I_{2,0,2}\approx \frac{8\pi^2\ln\frac{4 E^2 E'^2}{\sqrt{(m^2\omega^2- \lambda b_0 \omega E E')^2+\frac{16 E^4 E'^2}{\tau^2}}}}{3\left[(m^2\omega- \lambda b_0  E E')^2+\frac{16 E^3 E'}{\tau^2}\right]}\nonumber\\
    & \nonumber\\& \times \frac{\pi^2\tau^3}{4E^4E'^4  \omega}\left[E'^6\arctan\frac{m^2(\omega^*-\omega)\tau}{  E'(E-\omega^*)}+E^6\arctan\frac{m^2(\omega^*-\omega)\tau }{2 E (E-\omega^*)}\right]\Theta(\omega^*-\omega)\,,
\gal
where $\omega^*$ is given by \eq{f5}. The remaining integrals may be computed similarly, or related using the symmetry $p\rightarrow-p'$ such that
\ball{A8}
2\omega\kappa-k^2\rightarrow-(2\omega\kappa'+k^2)\,,\qquad 
\b q^2=-(p'-p+k)^2\rightarrow-(-p+p'+k)^2=\b q^2\,.
\gal

\bibliography{anom-biblio}

\end{document}